\documentclass[12pt]{article}
\usepackage{hyperref,amsfonts,amssymb,theorem,amsmath}

\newtheorem{statement}{Statement}

\newcommand{\HeunB}{\mathop{\rm HeunB}}
\newcommand{\HeunC}{\mathop{\rm HeunC}}

\title{Two new integrable cases of two-dimensional quantum mechanics with a magnetic field}
\author{V.G. Marikhin \footnote{e-mail: mvg@itp.ac.ru}\\
L.D. Landau Institute for Theoretical Physics, RAS\\
Chernogolovka, Moscow region, Russia}
\begin{document}
\maketitle
\thispagestyle{empty}
\begin{abstract}
Two integrable cases of  two-dimensional
Schr\"odinger equation with a magnetic field are proposed. Using the polar
coordinates and the symmetrical gauge, we will obtain solutions of these
equation through Biconfluent and Confluent Heun functions. The quantization
rules will be derived for both systems under
consideration.

\end{abstract}



\vspace{7mm}
The problem of obtaining integrable cases of the Schr\"odinger equation has a
long history. The number of such cases is not very large. The most famous
examples can be found in \cite{LLIII}, see also the celebrated work by Infeld
and Hull \cite{Infeld} and the classic papers by Schr\"odinger himself
\cite{Schrodinger_1940}--\cite{Schrodinger_1941b}. Among the recent studies we
can highlight  \cite{arm}--\cite{3r}. Most of these works deal with
one-dimensional problems.

The problem of integrable cases of the two-dimensional Schrodinger equation
with an electromagnetic field has long been studied as well, including the
celebrated ``Landau levels'' \cite{LLIII}, (see e.g. \cite{ferap} and the
references therein), but it is not completely solved yet. After the appearance
of modern methods for analyzing integrable systems, various aspects of
this problem were investigated. For instance, the class of (finite-gap)
solutions of the Schr\"odinger equation with a magnetic field was studied in
\cite{Dubr} (see also \cite{SP}). The factorization method was used in
\cite{ferap} to solve this problem (see also the references therein). The
Schr\"odinger equation in a magnetic field with additional linear and quadratic
integrals of motion were considered, for instance, in \cite{Berube} (also see
the references therein), where several interesting examples of such equations
in different coordinate systems were obtained.

The goal of this article is to introduce two new integrable examples of the
two-dimensional Schr\"odinger equation with an electromagnetic field. We
integrate these cases and describe the structure of the quantum states in terms
of the Biconfluent and Confluent Heun functions. We consider only the discrete
spectrum and the wave functions of the corresponding Schr\"odinger operator.

We need some properties of the above mentioned Biconfluent and Confluent Heun
functions, especially, the conditions when these functions reduce to
polynomials. One can find the necessary information in many works
\cite{Heun}--\cite{Plamen}, but for the reader's convenience and for further
applications, we re-obtain some of these properties in Appendix.

The author's work \cite{mvg1} contains a classification of all two-dimensional
Schr\"odinger operators with an additional integral of motion quadratic in the
momenta. Shortly speaking, one can derive new examples by starting from some
solution of the quasi-St\"ackel system (in St\"ackel coordinates) and
transforming this solution into the usual Cartesian coordinates. However, it is
a non-trivial problem to distinguish the examples which are interesting from
the physical standpoint, because of the large number of parameters contained in
the classification results. We postpone the comprehensive analysis of this
problem for the future work. At the moment, we present just two particular
examples with a reduced set of parameters. Namely, we consider the
Schr\"odinger operator in the usual form
\begin{equation}\label{H}
\hat{H}=\frac{1}{2m}\left(-i\hbar\vec{\nabla}-\frac{e}{c}\vec{A}\right)^2+u(r,\phi),\quad
\hat{H}\psi=E\psi,
\end{equation}
where the vector potential $A$ and the potential $u$ presented below contain
three independent parameters: the length $a$, a dimensionless parameter
$\epsilon$ and parameter $k$ which is integer in the case of the discrete
spectrum.

\section*{Example 1. Repulsive potential}
Let us consider the example
defined by the vector potential and potential
\begin{equation}\label{ex1}
A_{\phi}=\frac{c\hbar}{2e\,a^2}\,r\left(\epsilon+3\frac{r^2}{a^2}\right),\quad
A_r=0,\quad u=-\frac{\hbar^2}{2ma^2}\left(
2\frac{r^6}{a^6}+\epsilon\frac{r^4}{a^4}+2k\frac{r^2}{a^2}\right),
\end{equation}
and the energy
\begin{equation}\label{E1}
E=\frac{\hbar^2}{2ma^2}\lambda .
\end{equation}
We calculate a magnetic field
$$
B=\frac{c\hbar}{ea^2}(\epsilon+6\frac{r^2}{a^2}).
$$

At a first glance, the situation looks dangerous, because the potential
in ({\ref{ex1}) is strongly repulsive. However, it turns out that the
contribution of the magnetic field prevails the contribution of the potential.
A method to obtain an appropriate wave function is the following.

1. We substitute into (\ref{H}) the wave function of the form
\begin{equation}\label{psi1B}
\psi=e^{-\frac{r^4}{8a^4}}e^{-\epsilon\frac{r^2}{4a^2}}r^l e^{il\phi}w(r),
\end{equation}
where $l\geq 0$. The equation for $w$ is solved in terms the Biconfluent Heun
function
\begin{equation}\label{w1B}
w(r)=\HeunB(l,\epsilon,3l+2k,-\epsilon l-\lambda,\frac{r^2}{2a^2})
\end{equation}
with the parameters
$\alpha=l,\;\beta=\epsilon,\;\gamma=3l+2k,\delta=-l\epsilon-\lambda$, while its
second linearly independent solution is irregular in the point $r=0.$ The
requirement of the regularity of the $\HeunB$ function (\ref{w1B}) in the whole
range of definition implies that it must be a polynomial (see \cite{Andre} and
citations therein for more details). The polynomiality conditions are given in
Appendix.

The condition (\ref{uslB1}) in this case becomes $l+k=n+1.$ Thus, $k\leq n+1.$
We note that in this example $\det(A_{n+1})$ is a polynomial in $\lambda$ of
degree $n+1$ and the condition $\det(A_{n+1})=0$ defines $n+1$ values
$\lambda^i$ ($i=1,2,...,n+1$) and therefore $n+1$ energy levels $E^i$ by
formula (\ref{E1}).

To obtain the corresponding wave function, we substitute $\lambda=\lambda_i$
into the matrix $A_{n+1},$ compute $n+1$ eigenvectors
$\vec{p}^i=(p_0^i,p_1^i,..p_n^i)$ of this matrix and obtain the polynomials
$w^i(r)=P^i_n(\frac{r^2}{2a^2})$, where $P_{n}^i=\sum\limits_{j=0}^n p^i_jz^j.$
Finally, we substitute $w^i(r)$ into formula (\ref{psi1B}) in order to obtain
$n+1$ wave functions $\psi^i.$

At the moment, we have described the quantum states at fixed $n$ (or fixed
$l=n+1-k).$ In order to obtain the full picture, we have to fix the Hamiltonian
parameter $k,$ to consider all possible non-negative integers $l$ or
$n=k-1,k,k+1,k+2,...$ and to obtain an infinite set of ''blocks'' which contain $n+1$ energy levels and wave
functions for each permissible $n\geq k-1.$ Notice, that $i$ is an ``internal''
index for each $n.$

It is very important that only real roots $\lambda$ of equation (\ref{uslB2}) in this example are physical and belong to the discrete spectrum as well as corresponding
energies E and the wave functions $\psi.$ Therefore we must choose only these roots.

Thus, we obtain a rather complicated structure of the
quantum (discrete) states, which is non-typical for the quantum mechanics.
Such structure is similar for all cases under consideration.

All wave functions (\ref{psi1B}) under consideration are strongly localized in
the length scale of $a$ by the quadratic on $r$ magnetic field in this example.

We note that the diagonals of the matrix $A_{n+1}$ (\ref{matrix}) in this
example are defined by the following sequences:
\begin{equation}\label{abcB1}
a_j=\lambda-\epsilon (2j+1),\quad j=0,1,...,n,
\end{equation}
$$
b_j=2(j(j+n-k+3)+n-k+2),\quad j=0,1,...,n-1,
$$
$$
c_j=4(n-j),\quad j=0,1,...,n-1.
$$

2. The other possible substitution is
\begin{equation}\label{psi2B}
\psi=e^{-\frac{r^4}{8a^4}}e^{-\epsilon\frac{r^2}{4a^2}}r^l e^{-il\phi}w(r),
\end{equation}
where $l\geq 0$ as before. This gives us the following regular part of solution
\begin{equation}\label{w2B}
w(r)=\HeunB(l,\epsilon,-3l+2k,\epsilon l-\lambda,\frac{r^2}{2a^2}).
\end{equation}
Therefore, $\alpha=l,\;\beta=\epsilon,\;\gamma=-3l+2k,\delta=-\epsilon
l-\lambda.$ The condition (\ref{uslB1}) in this case becomes $-2l+k=n+1.$ Thus,
$k\geq n+1$, the Hamiltonian parameter $k$ must be a natural number and $k-n-1$
must be a non-negative even integer.

Subsequent consideration of this case is the same as in a previous case except
for the choice of the permissible $n:$ $n\geq 0,\; n=k-1,k-3,k-5,...$\,.
We choose only real $\lambda$ in this case as well as in previous one.

The diagonals of the matrix $A_{n+1}$ (\ref{matrix}) in this case are defined
by the sequences
\begin{equation}\label{abcB2}
a_j=\lambda-\epsilon (k-n+2j),\quad j=0,1,...,n,
\end{equation}
$$
b_j=j(2j-n+k+3)-n+k+1,\quad j=0,1,...,n-1,
$$
$$
c_j=4(n-j),\quad j=0,1,...,n-1.
$$

\section*{Example 2. Non-rational case}
This is defined by the vector potential and potential
\begin{equation}\label{ex2}
A_{\phi}=-\frac{c\hbar}{e}\,k\frac{a}{r\sqrt{r^2+a^2}},\quad A_r=0,\quad u=\frac{\hbar^2}{8ma^2}\left(\frac{3a^4}{(r^2+a^2)^2}-\epsilon\frac{a^2}{r^2+a^2}\right),
\end{equation}
and the energy of the form
\begin{equation}\label{E2}
E=-\frac{\hbar^2}{2ma^2}\chi^2.
\end{equation}
The magnetic field is equal to
$$
B=k\frac{c\hbar}{e}\frac{a}{(r^2+a^2)^{\frac{3}{2}}}.
$$
It is interesting that the full magnetic flux through the whole plane is
$\Phi=k\frac{2\pi\hbar c}{e}$, that is, it is an even multiple of the quantum
flux.

Let us use the change of variables
\begin{equation}\label{t}
t=\frac{1}{2a}(a+\sqrt{r^2+a^2}).
\end{equation}
We solve the equation (\ref{H}) with potentials (\ref{ex2}) in order to obtain
two linearly independent solutions which can be regular, but not
simultaneously, because the condition (\ref{uslC1}) differs in these cases. We
will consider both possibilities separately.

{\bf In the first case}, we obtain the wave function
\begin{equation}\label{psi1C}
\psi=e^{il\phi}\sqrt{2t-1}\,t^{\frac{k-l}{2}}(t-1)^{\frac{k+l}{2}}e^{2\chi t}\,\HeunC(4\chi,-l+k,k+l,0,\frac{1}{2}(k^2-l^2)+\frac{1}{4}(\epsilon+1)-\chi^2,t).
\end{equation}
Notice, that $k+l$ is a non-negative integer, because the solution must be
regular at the point $t=1$ ($r=0$). The condition (\ref{uslC1}) in this case
becomes $n+k+1=0.$ It means that $k$ is a negative integer. We substitute the
parameters of $\HeunC$ function as well as the necessary condition $n=-k-1$
into (\ref{abcC}) in order to obtain the diagonals of the matrix $A_{n+1}:$
\begin{equation}\label{abcC1}
a_j=\chi(\chi+2(2j-n-l))+l^2-n^2-n-j(j-2n-1)-\frac{1}{4}(1+\epsilon),\quad
j=0,1,...,n,
\end{equation}
$$
b_j=(j+1)(j-n-l),\quad j=0,1,...,n-1,
$$
$$
c_j=4(n-j)\chi,\quad
j=0,1,...,n-1.
$$

The function $\det(A_{n+1})$ is a polynomial in $\chi$ of degree $2(n+1)$ and
the condition $\det(A_{n+1})=0$ defines $2(n+1)$ values of $\chi.$ We must
choose only negative ones in order to obtain the regular (exponentially
decreasing with a distance) wave functions. Only quantum states with the negative $\chi$ belong to the discrete spectrum.

In this case we obtain the infinite set of matrices $A_{n+1}$ of the same size
$n+1=-k$, where the permissible $l$ are:
$$
l=-k,1-k,2-k,...
$$
The procedure to obtain the energy levels and wave function is the same as in
the previous example except for the definition of the energy levels which are
given by (\ref{E2}).

The scale of change of the wave function is $\frac{a}{|\chi|}$ and can be
arbitrary a priori.

{\bf In the second case}, we obtain the wave function in the form
\begin{equation}\label{psi2C}
\psi=e^{il\phi}\sqrt{2t-1}\,t^{\frac{l-k}{2}}(t-1)^{\frac{k+l}{2}}e^{2\chi t}\,\HeunC(4\chi,l-k,k+l,0,\frac{1}{2}(k^2-l^2)+\frac{1}{4}(\epsilon+1)-\chi^2,t).
\end{equation}
We note that $k+l$ is a non-negative integer, because the point $t=1$
($r=0$) must be regular, as before. The condition (\ref{uslC1}) in this case
takes the form $n+l+1=0.$ This means that $l$ must be a negative integer and therefore $k$ must be natural number. We
substitute the parameters of $\HeunC$ function as well as necessary condition
$l=-n-1$ into (\ref{abcC}) to obtain the diagonals of the matrix $A_{n+1}:$
\begin{equation}\label{abcC2}
a_j=\chi(\chi+2(2j-k-n))-j(j-2n-1)+n+\frac{1}{4}(3-\epsilon),\quad b_j=(j+1)(j-n-k),
\end{equation}
$$
c_j=4(n-j)\chi.
$$

In this case, we obtain a finite set of the matrices $A_{n+1}$ with the
permissible $n:$
$$
n\geq 0,\quad n=k-1,k-2,...,0.
$$

The procedure to obtain the energy levels (by
formula (\ref{E2}) in this case) and the wave function is same as in the
previous example, as well.

We must choose only negative $\chi$ to obtain the quantum states of the discrete spectrum as well as in the previous case.

We denote that values of the discrete spectrum and the corresponding wave functions can be obtained only numerically.
We postpone the numerical analysis of this problem for the future work.

\paragraph{Conclusion.} In this article, we have considered two absolutely
different integrable cases of two-dimensional quantum mechanics with the
electromagnetic field. Both cases reveal the rich and complicated structure of
the quantum states. These examples were integrated by use of the Confluent and
Biconfluent Heun functions. In our opinion, the Heun function and its four
confluent forms will be the main special functions of the 21st century. We
expect that many problems of quantum mechanics, as well as other applications,
will be resolved in terms of these transcendents. However, many results can be
obtained only numerically due to the complicated properties of the Heun
functions (e.g. calculation of large determinants and solution of the
corresponding eigenvalue problems, as in our case).

The approach proposed in this paper is based on the property of quantum
Liouville integrability. We hope that it can be applied for a wide range of
two-dimensional quantum mechanical problems.

\paragraph{Acknowledgments.} The author thanks V.E. Adler, I.V. Kolokolov,
and all participants of the mathematical physics seminar at the L.D. landau
Institute for Theoretical Physics for the useful discussions. This research was
supported in part by the Russian Foundation for Basic Research (Grant No.
16-01-00289).

\section*{Appendix}
\appendix
\section*{$\HeunB$ function}
The Biconfluent Heun function satisfies the following equation:
\begin{equation}\label{BHE}
\mbox{BHE}=y''(z)+\left(-2z-\beta+\frac{1+\alpha}{z}\right)y'(z)
 +\left(\gamma-\alpha-2-\frac{1}{2z}((1+\alpha)\beta+\delta)\right)y(z)=0.
\end{equation}
Its general solution is
\begin{equation}\label{BHS}
y(z)=C_1\HeunB(\alpha,\beta,\gamma,\delta,z)+C_2\HeunB(-\alpha,\beta,\gamma,\delta,z)z^{-\alpha}.
\end{equation}
Our aim here is to derive the necessary and sufficient conditions for the
function $\HeunB(\alpha,\beta,\gamma,\delta,z)$ to be polynomial. To do this in
a most convenient way, let us denote
\begin{equation}\label{polB}
\HeunB(\alpha,\beta,\gamma,\delta,z)=P_n(z)=\sum\limits_{j=0}^n p_jz^j,
\end{equation}
then we obtain the following recurrent relation for the coefficients $p_j:$
\begin{equation}\label{recB}
RB(j)=c_{j-1}p_{j-1}+a_jp_{j}+b_jp_{j+1}=0,
\end{equation}
where
\begin{equation}\label{abcB}
a_j=-(\delta+\beta(2j+\alpha+1)),\;
b_j=2(j(j+\alpha+2)+\alpha+1),\;
c_j=2(\gamma-\alpha-2j-2).
\end{equation}
We compute all $p_j$ starting from the initial values $p_{-1}=0,\;p_0=1$. Then,
the condition $b_{n+1}=0$ is equivalent to vanishing of the determinant of the
following three-ridiagonal matrix:
\begin{equation}\label{matrix}
A_{n+1} = \begin{pmatrix}
a_{0} & b_{0} & 0     & \cdots & 0 \\
c_{0} & a_{1} & b_{1} & \cdots & 0 \\
0     & c_{1} & a_{2} & \cdots & \vdots \\
\vdots & \vdots &\vdots& \ddots & b_{n-1} \\
0 & 0 & \cdots&c_{n-1} & a_{n}
\end{pmatrix}.
\end{equation}

The second necessary condition can be derived by the following trick. We
substitute $y(z)=z^n$ to (\ref{BHE}) and calculate the numerator (polynomial)
of the obtained expression. The vanishing of the leading term of this
polynomial implies the condition $\gamma-\alpha = 2(n+1).$ In this case, if the
matrix $A_{n+1}$ is degenerate then $b_{n+1}=0,\;b_{n+2}=0.$

Now, we can formulate the following statement.

\begin{statement}
The function $\HeunB(\alpha,\beta,\gamma,\delta,z)$ equals to a polynomial
$P_n(z)$ of degree $n$ if and only if the following two conditions are
satisfied:
\begin{equation}\label{uslB1}
\gamma-\alpha = 2(n+1),
\end{equation}
\begin{equation}\label{uslB2}
\det(A_{n+1})=0.
\end{equation}
\end{statement}

\section*{$\HeunC$ function}
The Confluent Heun function satisfies the equation
\begin{equation}\label{CHE}
\mbox{CHE}=y''(z)+\left(\alpha+\frac{\beta+1}{z}+\frac{\gamma+1}{z-1}\right)y'(z)
 +\left(\frac{\mu}{z}+\frac{\nu}{z-1}\right)y(z)=0,
\end{equation}
where
$\mu=\frac{1}{2}(\alpha-\beta-\gamma+\alpha\beta-\gamma\beta)-\eta,\quad
\nu=\frac{1}{2}(\alpha+\beta+\gamma+\gamma\alpha+\gamma\beta)+\delta+\eta$.
Then general solution of this equation is
\begin{equation}\label{CHS}
y(z)=C_1\HeunC(\alpha,\beta,\gamma,\delta,\eta,z)+C_2\HeunC(\alpha,-\beta,\gamma,\delta,\eta,z)z^{-\beta}.
\end{equation}

Let $\HeunC(\alpha,\beta,\gamma,\delta,z)=P_n(z),$
\begin{equation}\label{polC}
P_n(z)=\sum\limits_{j=0}^n p_jz^j,
\end{equation}
then we obtain the following recurrent relation for the coefficients $p_j:$
\begin{equation}\label{recC}
RC(j)=c_{j-1}p_{j-1}+a_jp_j+b_jp_{j+1}=0,
\end{equation}
where
\begin{equation}\label{abcC}
a_j=\mu-j(j-\alpha+\beta+\gamma+1),\quad
b_j=(j+1)(j+\beta+1),
\end{equation}
$$
c_j=(n-j)\alpha.
$$
The same scheme as in the previous section brings us to the following
statement.

\begin{statement}
The function $\HeunC(\alpha,\beta,\gamma,\delta,z)$ equals to a polynomial
$P_n(z)$ of degree $n$ if and only if the following two conditions are
satisfied:
\begin{equation}\label{uslC1}
\delta = -(n+1+\frac{1}{2}(\beta+\gamma))\alpha,
\end{equation}
\begin{equation}\label{uslC2}
\det(A_{n+1})=0.
\end{equation}
\end{statement}

\end{document}